\documentstyle[sprocl,bbox,epsf]{article}

\def\be{\begin{equation}}
\def\ee{\end{equation}}
\def\bea{\begin{eqnarray}}
\def\eea{\end{eqnarray}}
\begin{document}

\title{MULTIPION SYMMETRIZATION EFFECTS ON THE SOURCE DISTRIBUTION}

\author{Q. H. ZHANG}

\address{Institut f\"ur Theoretische Physik, Universit\"at Regensburg,
D-93040 Regensburg, Germany\\E-mail: qinghui.zhang@physik.uni-regensburg.de}


\maketitle\abstracts{ Any-order pion interferometry formulas for 
fixed pion multiplicity events and for mixed events are given. 
Multipion Bose-Einstein correlation effects on the two-pion 
interferometry and source distribution are studied. 
It is shown that generalized pion interferometry 
formula should depends on pion multiplicity distribution. 
Pion condensate is also discussed in the paper.}

\section{Introduction}
 Multipion Bose-Einstein(BE) correlations has now aroused 
great interests among 
physicists~\cite{Re,LL,Za,Pr,CGZ,ZCG,ZC,CZ,Z1,Z2,Z3,SL,M1,M2,BK,Wo,FWW,BZ,Ur,conf,ZSH,AL}. 
Among others, Lam and Lo~\cite{LL} were 
the first to suggest BASER concept in high energy physics. 
Zajc~\cite{Za} was the first to use Monte-Carlo methods to study multipion BE correlations 
effects on two-pion interferometry for fixed pion multiplicity event. 
Pratt~\cite{Pr} first suggested that multipion BE symmetrization effects 
may lead to BE condensate in high energy heavy-ion collisions. 
Detail derivation of multipion BE correlation effects on single 
particle spectrum and two-pion interferometry was given by 
Chao, Gao and I in Ref.~\cite{CGZ}. Recently Zim\'anyi and Cs\"org\H o suggested 
a new class density matrix and study multipion BE correlations 
and wavepacket effects on two-pion interferometry~\cite{ZC,CZ}. 
In Ref.~\cite{Z2}, I have 
derived any-order pion interferometry formula for a special class 
density matrix. In Ref.~\cite{Z3}, generalized any-order pion interferometry 
formulas are given which depends not only on the correlator as 
assumed previously but also on the pion multiplicity 
distribution which was neglected in the previous 
studies. Multipion BE correlation effects on the source 
distribution was studied in Ref.~\cite{ZSH}.
This paper is arranged as follows: 
In section 2, any-order pion interferometry 
formula for fixed pion multiplicity event are given. In section 3,
any-order pion interferometry formula for mixed events are derived.
In section 4, methods on multipion BE correlation simulation 
are discussed. 
Conclusions are given in section five.
 
\section{Any-order pion interferometry formulas for fixed pion 
multiplicity events}
Assume the source is totally chaotic. It is easily checked 
that the n-pion inclusive distribution in n-pion event can be 
expressed as~\cite{CGZ} 
\begin{equation}
P_n^{n}({\bf p_1,\cdot \cdot \cdot,p_n})=\sum_{\sigma}\prod_{j=1}^{n}\rho_{j,\sigma(j)}
\label{eq1}
\end{equation}
with
\begin{equation}
\rho_{ij}=\int d^4x g(x,\frac{p_i+p_j}{2})exp(i(p_i-p_j)\cdot x).
\end{equation}
Here $g(x,k)$ is a Wigner function which can be explained as the probability of fnding a pion 
at point $x$ with momentum $p$. 
$\sigma(j)$ denotes the $j$-th element of a permutations of the 
sequence $\{ 1,2,\cdot\cdot\cdot,n\}$, and the sum over $\sigma$ denotes 
the sum over all $n!$ permutations of this sequence.
Then the normalized $k$ pion inclusive distribution in 
$n$ pion events can be expressed as
\begin{equation}
P_n^{k}({\bf p_1,\cdot\cdot\cdot,p_k})=
\frac{\int P_n^n({\bf p_1,\cdot\cdot\cdot,p_n})\prod_{j=k+1}^{n}d{\bf p_j}}
{\int P_n^n({\bf p_1,\cdot\cdot\cdot,p_n})\prod_{j=1}^{n}d{\bf p_j}}.
\label{eq3}
\end{equation}
It is easily checked that $P_n^{k}({\bf p_1,\cdot\cdot\cdot,p_k})$ can be re-written as 
\begin{eqnarray}
&&P_n^k({\bf p_1,\cdot \cdot \cdot p_k})=
\frac{1}{n(n-1)\cdot\cdot\cdot (n-k+1)\omega(n)}
\sum_{i=k}^{n}\sum_{m_1=1}^{i-(k-1)}
\sum_{m_2=1}^{i-m_1-(k-2)}\cdot\cdot\cdot 
\nonumber\\
&&\sum_{m_{k-1}=1}^{i-m_1-m_2\cdot\cdot\cdot-m_{k-2}-1}
\sum_{\sigma}
G_{m_1}({\bf p}_1,{\bf p}_{\sigma(1)})
G_{m_2}({\bf p}_2,{\bf p}_{\sigma(2)})
\nonumber\\
&&
\cdot \cdot\cdot
G_{m_{k-1}}({\bf p}_{k-1},{\bf p}_{\sigma(k-1)})
G_{i-m_1\cdot\cdot\cdot-m_{k-1}}({\bf p}_{k},{\bf p}_{\sigma(k)})\omega(n-i)
\end{eqnarray}
with
\begin{equation}
\omega(n)=
\frac{1}{n}\sum_{i=1}^{n}\omega(n-i)\int d{\bf p}G_i({\bf p},{\bf p}),~~~\omega(0)=1
\end{equation} 
and
\begin{equation}
G_i({\bf p,q})=\int \rho({\bf p},{\bf p_1})
d{\bf p_1}\rho({\bf p_1},{\bf p_2})\cdot\cdot\cdot d{\bf p_{i-1}}\rho({\bf p_{i-1}},{\bf q}).
\end{equation}
Here $\sigma(i)$ denotes the $i$-th element of a permutations of the 
sequence $\{ 1,2,\cdot\cdot\cdot,k\}$, and the sum over $\sigma$ denotes 
the sum over all $k!$ permutations of this sequence.
Then the $k$-pion ($k\le n$) interferometry formula for n-pion evnets 
can be expressed as
\begin{equation}
C_n^{k}({\bf p_1,\cdot\cdot\cdot,p_k})
=\frac{P_n^{k}({\bf p_1,\cdot\cdot\cdot,p_k)}}{\prod_{j=1}^{k}P_n^{1}({\bf p_j)}}.
\end{equation}

We assume the source distribution as 
\begin{eqnarray}
g({\bf r},t,{\bf p})&=&
n_0\cdot (\frac{1}{2\pi R^2})^{3/2}exp(-\frac{{\bf r}^2}{2R^2})\delta(t)
(\frac{1}{2\pi \Delta^2})^{3/2} exp(-\frac{{\bf p}^2}{2 \Delta^2}).
\end{eqnarray} 
Here $n_0$ is a parameter. Then the multipion BE correlation effects on 
two-pion interferometry for fixed evnets are shown in Fig.1. It is 
easy to see that as pion multiplicity increases,  multipion BE correlations 
effects on the two-pion interferometry effects becomes stronger. In the 
 limit $n\rightarrow \infty$ we have $C_n^2=1$.
\begin{figure}[t]\epsfxsize=12cm \epsfysize=6cm
\centerline{\epsfbox{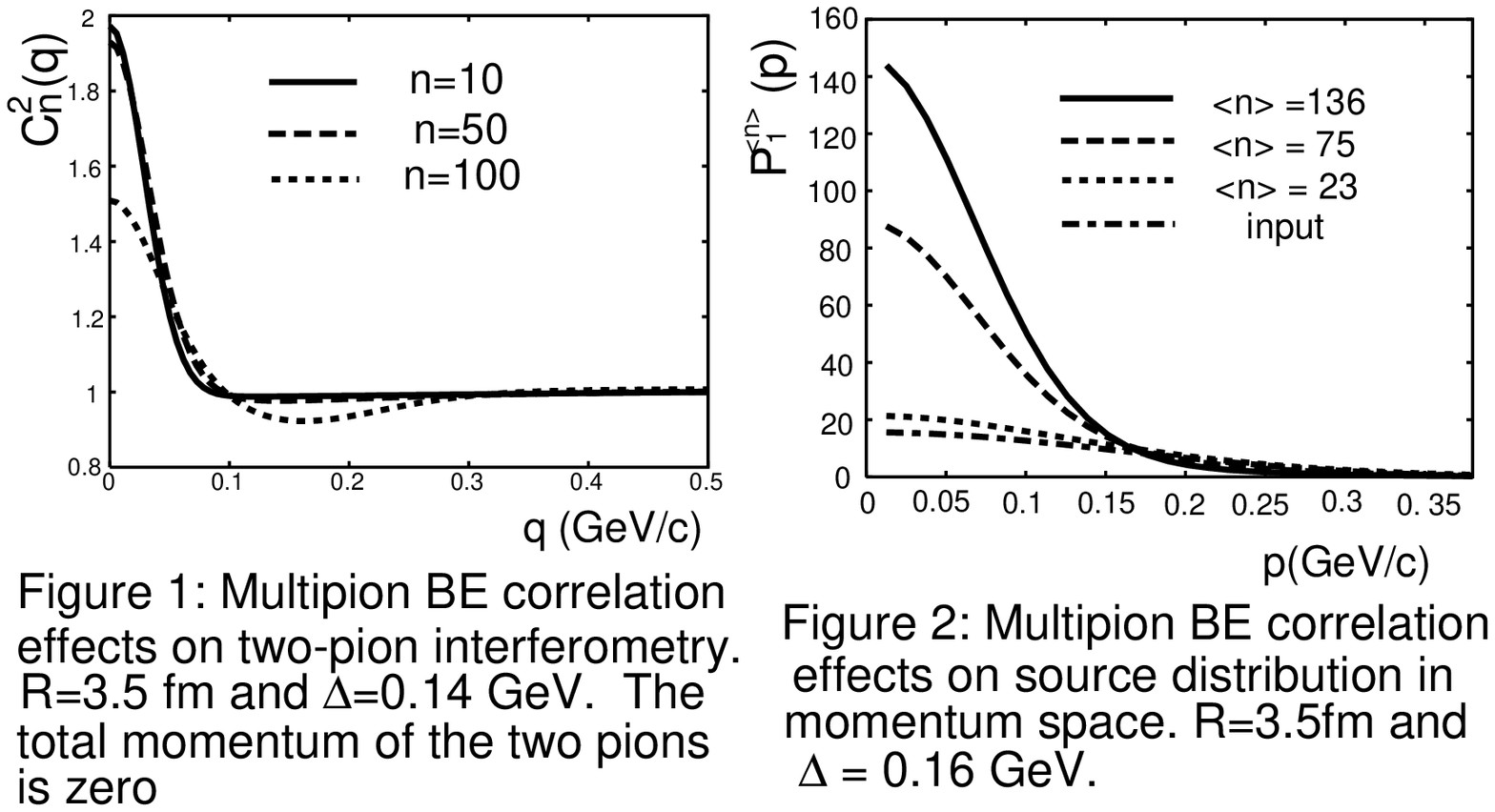}}
%\caption{\it multipion BE correlation effects on two-pion interferometry}
\vskip -0.5 cm
\end{figure}
%\vskip -6cm
\section{Any-order pion interferometry formula fo mixed events}

In the experimental analyses, one normally mixed all events to study 
pion interferometry. The $k$-pion inclusive distribution can be expressed as
\begin{equation}
N_k({\bf p_1,\cdot\cdot\cdot,p_k})=\sum_{n=k}^{\infty} 
P_n \cdot n\cdot(n-1)\cdot\cdot\cdot (n-k+1)P_n^k({\bf p_1,\cdot\cdot\cdot,p_k}).
\end{equation}
Here $P_n$ is the normalized pion multiplicity distribution. It is easily checked that 
\begin{equation}
\int N_k({\bf p_1,\cdot\cdot\cdot,p_k})\prod_{j=1}^{k}d{\bf p_j}=
\langle n(n-1)\cdot\cdot\cdot (n-k+1)\rangle~~~ ,
\end{equation}
then the $k$-pion interferometry formulas can be expressed as
\begin{equation}
C_k({\bf p_1,\cdot\cdot\cdot,p_k})=
\frac{N_k({\bf p_1,\cdot\cdot\cdot,p_k})}{\prod_{j=1}^{k}N_1({\bf p_j})}.
\end{equation}
The discussion about the definition of two-pion interferometry formulas can be found in 
the paper by Miskowiec and Voloshin~\cite{MV} and can also be found in Ref.~\cite{ZSH}. Using 
Eq.(4) , 
the $k$-pion inclusive distribution can be expressed as~\cite{Z3}
\begin{eqnarray}
&&N_k({\bf p_1,\cdot \cdot \cdot p_k})=
\sum_{n=k}^{\infty}P_n\frac{\omega(n-i)}{\omega(n)} \sum_{i=k}^{n}\sum_{m_1=1}^{i-(k-1)}
%\sum_{m_2=1}^{i-m_1-(k-2)}
\cdot\cdot\cdot 
\sum_{m_{k-1}=1}^{i-m_1-m_2\cdot\cdot\cdot-m_{k-2}-1}
\sum_{\sigma}
\nonumber\\
&&
G_{m_1}({\bf p}_1,{\bf p}_{\sigma(1)})
%G_{m_2}({\bf p}_2,{\bf p}_{\sigma(2)})
\cdot \cdot\cdot
G_{m_{k-1}}({\bf p}_{k-1},{\bf p}_{\sigma(k-1)})
G_{i-m_1\cdot\cdot\cdot-m_{k-1}}({\bf p}_{k},{\bf p}_{\sigma(k)})
\nonumber\\
&&=
\sum_{m_1=1}^{\infty}\cdot\cdot\cdot 
\sum_{m_k=1}^{\infty}
h_{m_1+\cdot\cdot\cdot+m_k}
\sum_{\sigma}
G_{m_1}({\bf p}_1,{\bf p}_{\sigma(1)})
\cdot \cdot\cdot
G_{m_{k}}({\bf p}_{k},{\bf p}_{\sigma(k)})
\end{eqnarray}
with
\begin{equation}
h_{m_1+\cdot\cdot\cdot+m_k}=\sum_{n=m_1+\cdot\cdot\cdot+m_k}^{\infty}P_n
\frac{\omega(n-m_1-\cdot\cdot\cdot-m_k)}{\omega(n)}.
\end{equation}

It is interesting to notice that if 
$P_n=\omega(n)/\sum_n \omega(n)$ as assumed in Ref.~\cite{Z2,ZC,Pr}, we have 
the following modified $k$ pion inclusive distribution~\cite{Z2}:
\begin{equation}
N_k({\bf p_1,\cdot \cdot \cdot p_k})=
\sum_{\sigma}H_{1\sigma(1)}\cdot\cdot\cdot H_{k\sigma(k)}
\end{equation}
with
\begin{equation}
H_{ij}=H({\bf p_i,p_j})=\sum_{n=1}^{\infty}G_n({\bf p_i,p_j}).
\end{equation}
One interesting property about Eq.(15) is: Eq.(15) is 
very similar to the pure $n$ pion inclusive distribution (Eq.(1)) but with a modified source 
distribution $S(x,K)$ which satisfies
\begin{equation}
H({\bf p_i,p_j})=\int S(x,\frac{p_i+p_j}{2})\exp(i(p_i-p_j)\cdot x)d^4 x.
\end{equation}
Based on Eq.(16), we will study multipion BE correlations effects
on the source distributions~\cite{ZSH}. From Eq.(16), the single 
particle spectrum distribution 
,$P^{\langle n\rangle}_1({\bf p})$, 
can be expressed as:
\begin{equation}
P^{\langle n\rangle}_1({\bf p})=\frac{N_1({\bf p})}{\langle n\rangle}
=\frac{\int S(x,p)d^4 x}{\langle n\rangle},~~
\langle n \rangle =\int N_1({\bf p})d{\bf p}=\int S(x,{\bf p})d^4x d{\bf p}.
\end{equation}
Similarly the source distribution in coordinate space 
,$P^{\langle n\rangle}_1({\bf r})$, 
can be expressed as
\begin{equation}
P^{\langle n\rangle}_1({\bf r})
=\frac{\int S(x,{\bf p})d{\bf p} dt}{\langle n\rangle}.
\end{equation}
Assume the input $g(x,p)$ as Eq.(8), multipion BE correlation effects on 
the source distribution are shown 
in Fig.2 and Fig.3. It is clear that as $\langle n\rangle$ increases, multipion BE 
correlations effects on the source distribution becomes stronger. Multipion 
BE correlation make pions concentrate in momentum and coordinate space. 
It is easily checked that the half width of the source distribution 
in momentum space and coordinate 
space $\Delta_{eff}$ and $R_{eff}$ satisfies the following relationship:
\begin{equation}
\sqrt{\frac{\Delta}{2R}}\le \Delta_{eff}\le \Delta,~~~
\sqrt{\frac{R}{2\Delta}}\le R_{eff}\le R.
\end{equation}
If we take $\frac{\langle n \rangle}{(R\Delta)^3}\rightarrow \infty$. 
we have $R_{eff}=\sqrt{\frac{R}{2\Delta}}$ and 
$\Delta_{eff}=\sqrt{\frac{\Delta}{2R}}$. The source size satisfy the minimal 
Heisenberg relationship. That is all pions are concentrated in a single 
phase space and pion condensate occurs. 
But the pion multiplicity distribution is Boson form:$P_n=\frac{\langle n\rangle^n}
{(\langle n\rangle +1)^{n+1}}$. 

In general, pion interferometry formula depends not only on the correlator 
$\rho_{i,j}$ but also on $P_n$ as shown in Eq.(12). The later 
property was neglected in the previous studies. 
Recently, Egger, Lipa and Buschbeck~\cite{ELB} 
results seems support the conclusion presented here. Previous studies based 
on pure multipion interferometry formulas  
have shown that the difference between three-pion interferometry and two-pion 
interferometry are not so larger~\cite{HZ97,conf1}. Certainly, there is also 
another possibility that the there are some correlation among the 
emitted pions which was not include in the above derivation. 
So one needs more studies about the two-pion and higher-order pion 
interferometry!
 
If we assume the pion state
as
\begin{equation}
|\phi\rangle =\sum_n a_n |n>.
\end{equation}
Here $a_n$ is a parameter which connected with pion multiplicity distribution.
$|n>$ is the non-normalized pion $n$ pion state which can be expressed as
\begin{equation}
|n\rangle =\frac{(\int d{\bf p} j({\bf p}) a^+({\bf p}))^n}{n!}|0>.
\end{equation}
Here $a^+({\bf p})$ is the pion creation operator and $j({\bf p})$ is the pion 
probability amplitude. 
Then the pion multiplicity distribution $P_n$ can be expressed as
\begin{equation}
P_n=\frac{|a_n|^2\omega(n)}{\sum_n |a_n|^2\omega(n)}.
\end{equation}
The two-pion interferometry results for different $P_n$(different $a_n$)is 
shown in Fig.4. One can see clearly that there are big differences  
among the two-pion interferometry results for different $P_n$. 

\section{Monte-Carlo simulation of multipion interferometry}

To put two-pion BE correlations in event generator has been studied by 
different authors~\cite{CGZP}. To put multipion BE correlations in event generator is 
a very difficult task as shown in Ref.~\cite{Za,FWW}. 
Recently Fialkowski et al~\cite{FWW} used the 
methods of Bialas, Krzywicki~\cite{BK} and Wosiek~\cite{Wo} and implemented 
approximately multipion BE 
correlation in JETSET/PYTHIA to study multipion BE correlation effects on the W mass. 
Due to the fact that there is no space-time picture in the model 
JETSET, So they had to put $\rho_{ij}$ by hand. On the other 
hand for model which has space-time picture, one can construct 
$g(x,p)$ according to following method~\cite{ZZZ,ZC}: 
$D(x,p)$ is a total classical function which is the output of the evnet generator 
 and can be written down as:
\begin{equation}
D(x,p)=\sum_i \delta(x-x(i))\delta(p-p(i)). 
\end{equation}
Here $x(i)$ and $p(i)$ are the $i$ particle coordinate and momentum. 
The natural and easiest way to construct $g(x,p)$ the semiclassical 
function  is to replace the 
above delta function with a Gaussian wavepacket 
\begin{eqnarray}
g(y,k)&=&\sum_i \prod_{j=0}^{3} \exp(
-\frac{(y_j-x(i)_j)^2}{\delta x_j^2})
\exp(-\frac{k_j-p(i)_j}{\delta p_j^2}
%-\frac{k_1-p(i)_1}{\delta p_1^2}
%-\frac{k_2-p(i)_2}{\delta p_2^2}
%-\frac{k_3-p(i)_3}{\delta p_3^2}
).
\end{eqnarray}
In general $\delta x_j$ ($\delta p_j$) should not be the same. But for simplicity one can 
set them the same value and take $\delta p_j=\frac{1}{\delta x_j}$. 
The same discussion can be found 
in the Ref.~\cite{PGG90}. 
Unfortunately the 
calculating work for the above methods to put multipion BE 
correlation in event generator is so large that a new  
method which enable us to implement quickly  multipion correlation 
in event generator is a debate for physics in this field. 
\begin{figure}[t]\epsfxsize=12cm \epsfysize=6cm
\centerline{\epsfbox{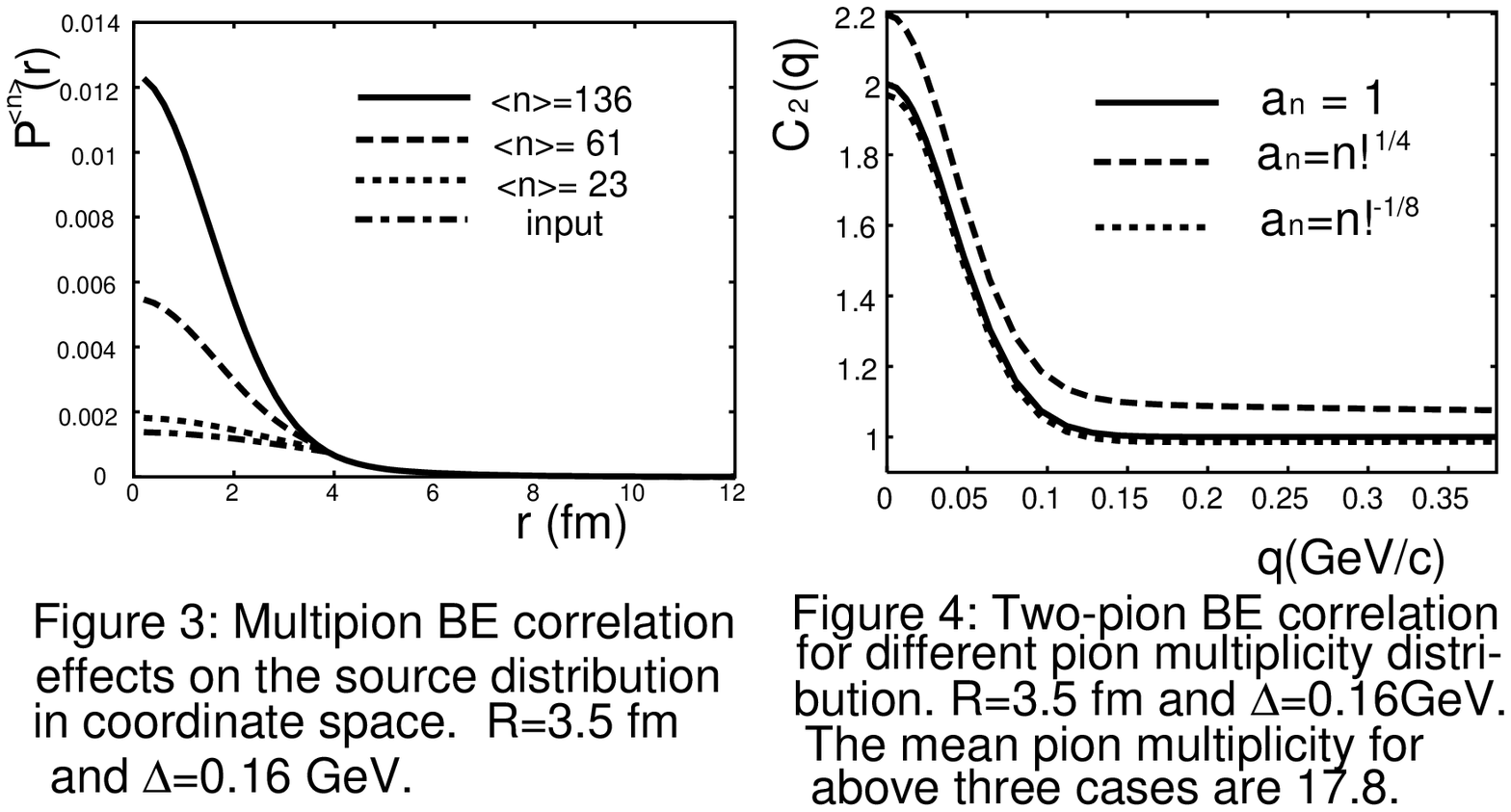}}
%\caption{\it multipion BE correlation effects on two-pion interferometry}
\vskip -0.5cm
\end{figure}
%\vskip -6cm
\section{Conclusions}
In the above, we mention nothing about the effects of 
flow~\cite{xxx} and resonance~\cite{yyy} 
on the pion interferometry. As our main interests is the 
BE statistical effects and the flow and resonance 
 can be included in the function $g(x,p)$. 
Certainly the multipion coulomb effects is 
still a open question in this field. Energy constraint effects 
on the pion interferometry can be studied by the method 
presented in Ref.~\cite{ZCG}.  In this paper, we have derived 
any-order pion interferometry formula for fixed multipion 
events and for mixed events. In general pion interferometry formulas
should depends on not only on the correlator but also 
on the pion multiplicity distribution which was neglected 
in previous studies. It is shown that as the pion 
multiplicity becomes very larger all pions concentrate in 
a single phase space and pion condensate 
occurs. But the pion multiplicity distribution is BE form. 
The methods to put multipion BE correlations in the 
event generator is discussed.

%\begin{figure}[t]
%\rule{5cm}{0.2mm}\hfill\rule{5cm}{0.2mm}
%\vskip 2.5cm
%\rule{5cm}{0.2mm}\hfill\rule{5cm}{0.2mm}
%\psfig{figure=ha1.eps,height=1.5in}
%\caption{A generalized cactus tree: the confluent
%transfer-matrix $S$ transforms the state function $f(x)$ and
%$f(z)$ into $f(x)$.  \label{fig:radish}}
%\end{figure}
%\begin{figure}[t]\epsfxsize=11cm
%\centerline{\epsfbox{ha11.eps}}
%\caption{\it multipion BE correlation effects on two-pion interferometry}
%\end{figure}
%\rule{5cm}{0.2mm}\hfill\rule{5cm}{0.2mm}
%\vskip 2.5cm
%\rule{5cm}{0.2mm}\hfill\rule{5cm}{0.2mm}
%\psfig{figure=ha1.eps,height=1.5in}
%\caption{A generalized cactus tree: the confluent
%transfer-matrix $S$ transforms the state function $f(x)$ and
%$f(z)$ into $f(x)$.  \label{fig:radish}}
%\end{figure}

\section*{Acknowledgments}

Q.H.Z. thanks Drs. U. Heinz, J. Zim\'anyi, 
T. Cs\"org\H o, R. Lednicky, Y. Sinyukov, D. Mi\'skowiec, H. Egger, 
B. Buschbeck, P. Scotto and Urs. Wiedemann 
for helpful discussions. 
Part of the talk is based on the work with U. Heinz and P. Scotto.
Q.H.Z was supported by the Alexander 
von Humboldt Foundation.

\end{document}